# Three Dimensional Radiation Hydrodynamic Simulations of Massive Star Envelopes

Yan-Fei Jiang (姜燕飞)[1], Matteo Cantiello[2,3], Lars Bildsten[1,4], Eliot Quataert[5], Omer Blaes[4], James Stone[3]

[1]Kavli Institute for Theoretical Physics, University of California, Santa Barbara, CA 93106, USA

[2]Center for Computational Astrophysics, Flatiron Institute, 162 Fifth Avenue, New York, NY 10010, USA

[3]Department of Astrophysical Sciences, Princeton University, Princeton, NJ 08544, USA

[4]Department of Physics, University of California, Santa Barbara, CA 93106, USA

[5]Astronomy Department and Theoretical Astrophysics Center, University of California at Berkeley, Berkeley, CA 94720-3411, USA

**Stars more massive than $20-30M_\odot$ are so luminous that the radiation force on the cooler, more opaque outer layers can balance or exceed the force of gravity. When exactly balanced, the star is said to be at the Eddington limit[1]. These near or super-Eddington outer envelopes represent a long standing challenge for calculating the evolution of massive stars in one dimension[2], a situation that limits our understanding of the stellar progenitors of some of the most exciting and energetic explosions in the universe. In particular, the proximity to the Eddington limit has been the suspected cause for the variability, large mass loss rate and giant eruptions of an enigmatic class of massive stars: the luminous blue variables (LBVs) [3–7]. When in quiescence, LBVs are usually found on the hot ($T_{\rm eff}\simeq 2-4\times 10^4$K) S Dor** **instability strip. While in outburst, most LBVs stay on the cold S Dor instability strip with a $T_{\rm eff} \simeq 9000{\rm K}$[5,8]. Here we show that physically realistic three dimensional global radiation hydrodynamic simulations of radiation dominated massive stars with the largest supercomputers in the world naturally reproduce many observed properties of LBVs, specifically their location in the Hertzsprung-Russell (HR) diagram and their episodic mass loss with rates of $10^{-7} - 10^{-5} M_\odot/yr$. The helium opacity peak is found to play an important role to determine these properties, which is not realized in the traditional one dimensional models of massive stars. The simulations also predict that convection causes irregular envelope oscillations and 10-30% brightness variations on a typical timescale of a few days. The variability is more prominent in our models that are on the cool part of the S Dor instability. These calculations pave the way to a quantitative understanding of the structure, stability and the dominant mode of mass loss of massive stars.**

Typical LBVs have luminosities $6 \times 10^5$ to $\sim 4 \times 10^6$ times the solar luminosity, and effective temperatures either hotter than $\sim 2 \times 10^4 {\rm K}$ in quiescence or around 9000K. As shown in Figure 1, these temperatures are nearly independent of the luminosity during outburst. We use the 1D stellar evolution code MESA[2,9–11] to evolve solar metallicity stars with initial mass $M_i$ = 35 and $80 M_\odot$ to this region of the Hertzsprung-Russell (HR) diagram[12]. These 1D evolution models for stars with luminosities close to the Eddington limit are very uncertain and different groups[13–15] get significantly different results as illustrated in Figure 1. However, they do provide a useful first approximation of the physical conditions in the radiation dominated envelopes of these stars. One important feature of these regions is the presence of an opacity peak at temperatures $T \simeq 1.8 \times 10^5$ K due to lines of iron-group elements, which is usually called the iron opacity peak. The opacity there is often a factor of a few larger than that from free electron scattering and can cause the local radiation acceleration to exceed that from gravity[2]. In this situation, the envelope is unstable to convection at the location of the iron opacity peak[2,16], with the properties of convection and envelope structure depending crucially on just how deep within the star this iron opacity peak occurs[12]. The opacity can, in principle, increase further to values 100 times that of electron scattering when the temperature drops below $1-4\times 10^4$K. This arises from helium and hydrogen recombination, but only as long as the density exceeds $\sim 10^{-9}$ g/cm$^3$, a value much larger than what is realized in 1D hydrostatic structures around this temperature range.

We take the typical luminosity, density and gravity at the location of the iron opacity peak in our 1D models and construct envelopes in hydrostatic and thermal equilibrium in spherical polar geometry covering the temperature range from $10^4 - 10^6$K as the initial conditions for our 3D simulations. We studied three simulations: a model corresponding to a star with $\log(L/L_\odot) = 6.2$ and $T_{\rm eff} = 9000$K (T9L6.2), one with $\log(L/L_\odot) = 6.4$ and $T_{\rm eff} = 19000$K (T19L6.4) and one with $\log(L/L_\odot) = 6.0$ and $T_{\rm eff} = 19000$K (T19L6). Although the three runs are based on 1D models for stars with different initial masses and/or at different evolutionary stages, they share similar properties such as density and the Eddington ratio, which is the ratio between radiation and gravitational accelerations, at the iron opacity peak. The main difference in these models is the pressure scale height at the iron opacity peak, which results in different total mass and optical depth above the convective region and hence different surface temperature[12].

We used 60 millions CPU hours in the supercomputer Mira awarded by Argonne Leadership Computing Facility for the INCITE program, as well as computational resources from NASA and NERSC, to solve the 3D radiation hydrodynamic equations[17] and follow the physically realistic evolution of these envelopes. These simulations take the fixed core mass and luminosity coming from the bottom boundary as input and then determine the envelope structure, effective temperature and mass loss rate self-consistently. The bottom boundary has a density and temperature order of magnitudes larger than the values at the iron opacity peak and remains relatively still. Both OPAL Rosseland and Planck mean opacity tables[18] are included in the simulations to capture the momentum and thermal coupling between the radiation field and gas[19] . The histories of azimuthally averaged radial profiles of density, turbulent velocity, radiation temperature and opacity for the run `T9L6.2` are shown in Figure 2. The envelope is convectively unstable at the iron opacity peak because the local radiation acceleration is larger than the gravitational acceleration [2, 12, 20], which causes the density to increase with radius around that region in the initial hydrostatic structure as shown in the top panel of Figure 2. Convection takes about 10 dynamical times ($\approx 43$ hours) to destroy the density inversion, which causes high density clumps to rise, expand and cool. When the temperature drops below $\approx 6 \times 10^4$K with a much higher density compared with the density at that temperature prior to the onset of convection, a strong helium opacity peak appears (bottom panel of Figure 2). Since the local radiation acceleration is now 10 times larger than the gravitational acceleration, a large fraction of the envelope expands dramatically, with most of the gas above that region blown away with an instantaneous mass loss rate $\approx 0.05 M_\odot$/yr. After 400 hours, the envelope settles down to a steady-state structure, as shown in the right panels of Figure 2. Convection is still operating around $80-90R_\odot$, with a second helium opacity peak around $200R_\odot$. Convection also causes envelope oscillations with a typical time scale of a day. The time averaged location of the photosphere, where the total Rosseland optical depth to the outer boundary of the simulation box is 1, is at $342.8R_\odot$ as indicated by the dashed blue lines in Figure 2 with an averaged radiation temperature $9.06\times 10^3$K at that location. The mass average turbulent velocity is only $1\%$ of the sound speed deep in the envelope, but it becomes supersonic near the photosphere, causing strong shocks and large temperature and density fluctuations near the photosphere. During each oscillation cycle, as indicated by both the density and turbulent velocity in the right panel of Figure 2, part of the mass becomes unbound with mass loss rate $\approx 5\times 10^{-6} M_\odot$/yr. This simulation naturally produces a massive star with luminosity, effective temperature and mass loss consistent with LBVs when they are on the cold, $T_{\rm eff}=9000$K, S Dor instability strip during an outburst.

The run `T19L6.4` shows a similar evolution history. Convection develops around the iron opacity peak and destroys the initial density inversion. After removing the initial gas above that region, a steady state structure is formed. One snapshot of the 3D density and radiation energy density structures is shown in Figure 2. Due to a smaller pressure scale height and a smaller optical depth across the typical convective element, the gas rising due to convection experiences a much smaller temperature change. This causes a much lower value of the opacity at the helium peak compared with the run `T9L6.2`, and thus a smaller total optical depth above the iron opacity peak region. Although the luminosity for this run is a little bit larger than the previous model, the less significant helium opacity peak places the time averaged location of the photosphere at a smaller radius $102R_\odot$ with a higher effective temperature $1.87\times 10^4$K, which confirms that without the helium opacity peak, the star will not move to the cold S Dor instability strip. The presence of a smaller helium opacity peak results in a substantial reduction of the envelope oscillation amplitude and a lower associated mass loss rate of $\approx 1 \times 10^{-6} M_\odot/\text{yr}$.

The final run T19L6 has very similar properties to T19L6.4, in particular a comparable value of the pressure scale height at the iron opacity peak. However, the model is calculated for a smaller core mass and a smaller luminosity. In steady state the envelope solution has a $T_{\text{eff}} = 1.89 \times 10^4 \text{K}$ with time averaged photosphere radius of $63.7 R_\odot$, and an episodic mass loss rate associated with envelope oscillations of only $\approx 5 \times 10^{-7} M_\odot/\text{yr}$. This confirms that when the iron opacity peak is found in a region with a small pressure scale height (mainly due to a larger gravitational acceleration at that location), the effective temperature remains too hot for the helium opacity to become significant, and the massive star model stays closer to the hot side of the S Dor instability strip. When the pressure scale height increases (for example, as the star expands due to evolution), a significant helium opacity peak will appear and force the star to the cold S Dor instability strip. At the same time, the mass loss rate will also increase by more than a factor of 5, which are all properties consistent with the behavior seen when an LBV goes into outburst.

Our simulations predict that LBVs in outburst should show irregular variability with typical timescales of days. In particular, we expect the variability pattern to be different for massive stars on the hot and cold S Dor instability strips, as shown in Figure 4. For massive stars with effective temperature near $9 \times 10^3 \text{K}$, a significant helium opacity peak exists in the envelope and causes large amplitude oscillations. The predicted stellar brightness then varies by a factor of $\approx 1.5 - 2$ in a day (consistent with the dynamical time scale at the iron opacity peak) as shown in the top panel of Figure 4. For stars with hotter effective temperatures near $1.9\times 10^4$K, for which the envelope helium opacity peak is much weaker, the variability has a much smaller amplitude. However, the luminosity can still vary by $\approx 20\%$ on timescales of a week to a few weeks, corresponding to the thermal time scale of the envelope above the iron opacity peak. This kind of variability has been seen in recent high cadence observations of massive stars [21–23], and the correlation between variability and effective temperature can be tested with future observations. The envelope is loosely-bound and dominated by turbulent convection (Figure 2), so the oscillation at the stellar surface is chaotic. However, there are moments in the envelope evolution when the majority of the photosphere is falling back onto the core, as suggested by the integrated luminosity shown in Figure 3. This can potentially explain the time dependent behavior of P-Cygni and inverse P-Cygni profiles found in some LBVs[21,23,24].

The mass loss rates we obtain from our steady-state simulations are broadly consistent with the inferred mass loss rate during both the quiescent and outburst phases of LBVs [7,23]. We also naturally get a larger mass loss rate during outburst compared with the value during quiescence. We find that the physical mechanism responsible for driving the dominant mode of mass loss in LBV stars is the interaction of their large radiative flux with opacity peaks that appear in the optically thick envelope of these stars as they expand and cool. Traditionally, mass loss due to radiation force on the ultraviolet lines in the optically thin region is thought to be the dominant mechanism for these massive stars[25,26]. Our work shows that continuum radiation alone in the optically thick envelope, where the classical line driven wind theory does not apply, can already drive mass loss rate comparable to the observed value. Importantly, while the iron opacity peak is strongly metallicity dependent, as long as a turbulent stellar envelope cools to low-enough temperatures, the helium and hydrogen opacity peaks will always cause large Eddington factors. This suggests that this mode of mass loss may be less sensitive to metallicity than line-driven winds.

The simulations also suggest possible paths for massive stars to transition between the hot and cold S Dor instability strips as indicated by the dotted black lines for the observed LBVs in Figure 1. Starting from the hot S Dor instability strip, stars will expand due to nuclear evolution and the pressure scale height at the iron opacity peak region will increase. When a significant helium opacity peak appears, the star will undergo outburst and move to the cold S Dor instability strip. The amount of mass initially above the iron opacity peak region can only sustain the mass loss rate for $\sim 10$ years, while the thermal time scale below the convective region in the envelope is also comparable to $\sim 10$ years, which means the star may lose a significant fraction of the mass above the iron opacity peak via the wind before it has time to adjust to a new structure to keep this large mass loss rate. This will reduce the total optical depth above the iron opacity peak and increase the effective temperature. When the helium opacity peak is significantly reduced, these stars will return to the hot S Dor instability strip on this time scale. As the iron opacity peak moves to the deeper region of the envelope due to stellar evolution, this process can repeat. Alternatively, if the massive star is in a binary system as suggested for some LBVs [27] and the companion deposits mass on the surface of the star, the additional mass will likely be ejected by the massive star as we found in our initial evolutions for each numerical simulation. This can be a trigger of the giant eruption of some LBVs. Detailed properties of this process will need to be studied with future calculations.

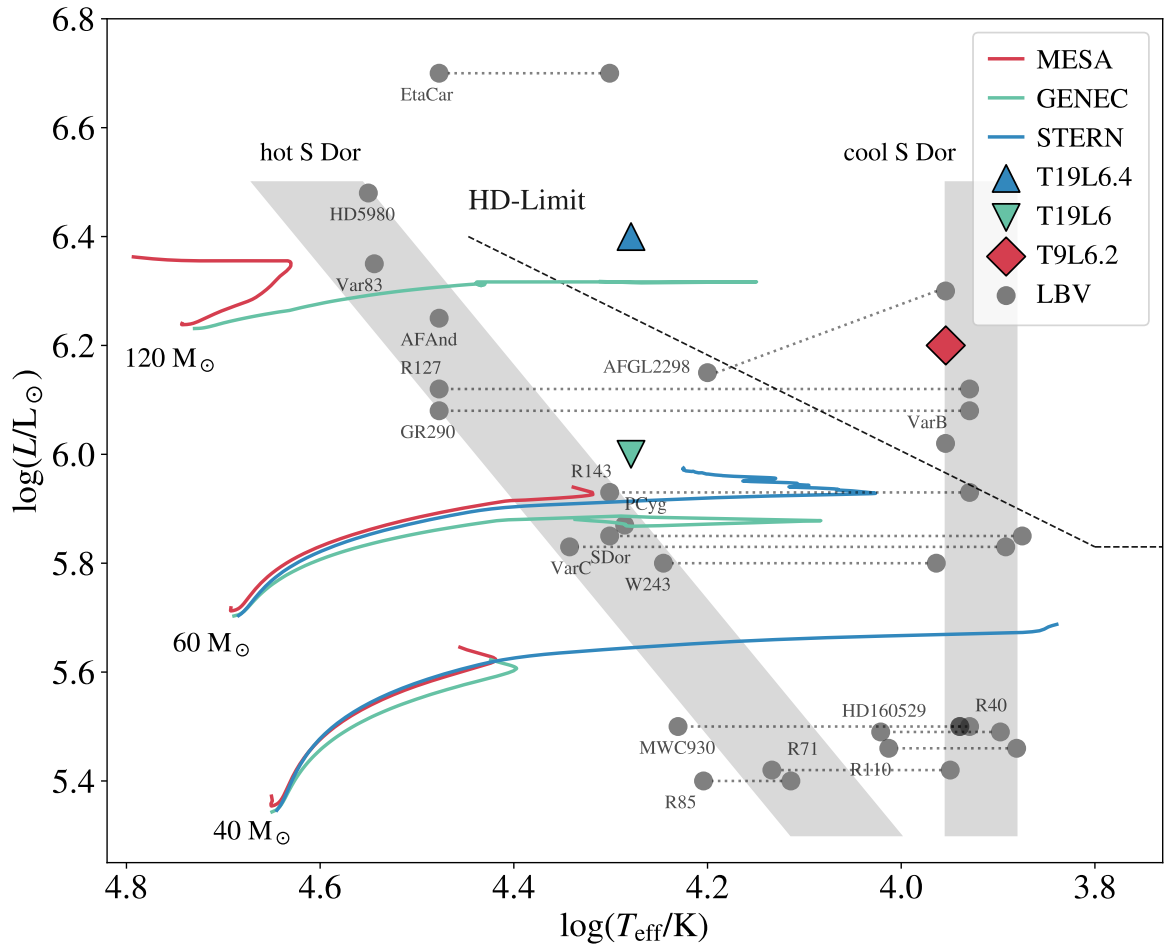

**Figure 1** | Hertzsprung-Russell diagram for luminous blue variables (LBVs, solid black dots). Shaded areas represent the locations where LBVs are most commonly found[8,28]: The diagonal band is the hot S Dor instability strip (LBVs in quiescence), while the vertical shaded band is the cool S Dor instability strip (LBVs in outburst). Dotted lines show observed excursions from quiescence to outburst. The solid red, green and blue lines correspond to main sequence, 1D stellar evolution tracks with different initial masses as calculated by different groups [13–15]. The locations of our three simulated stars are indicated by colored polygons. The dashed black line is

the Humphreys-Davidson (HD) limit[3].

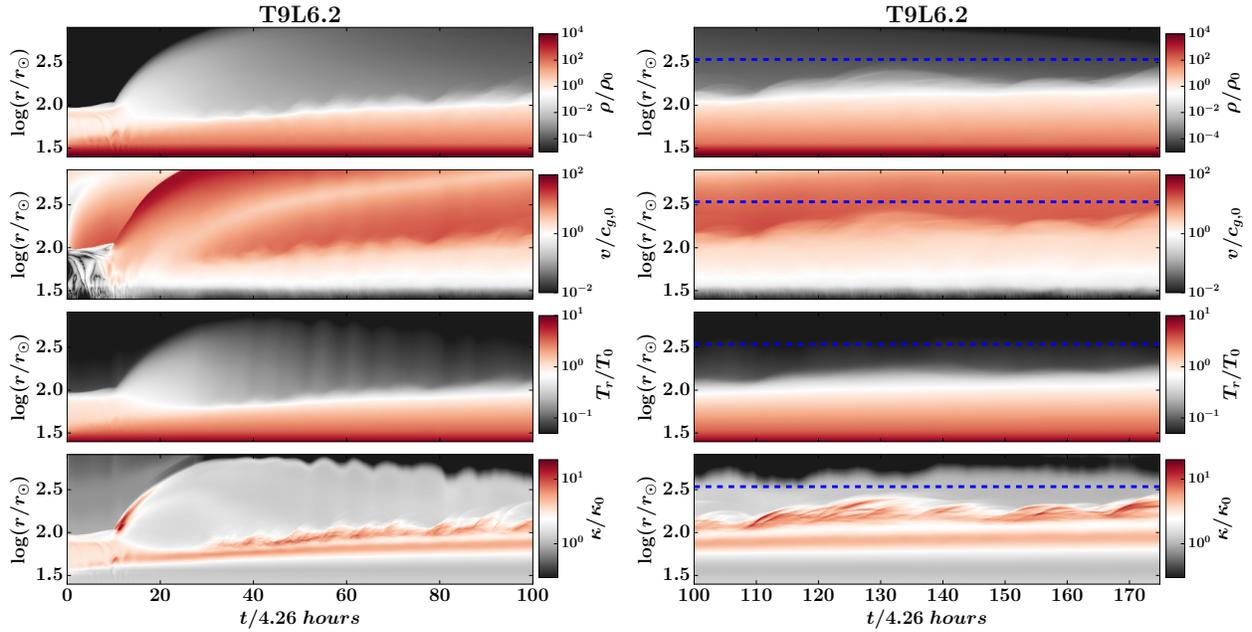

**Figure 2** | History of the azimuthally averaged radial profiles for the run T9L6.2. The left and right panels break at $t = 426$ hours to separate the initial transition and the steady state structures. From top to bottom, the four panels are for density $\rho$ (in unit of $\rho_0 = 3.6 \times 10^{-9}$ g/cm$^3$), turbulent flow velocity $v$ (in units of the isothermal sound speed near the iron opacity peak $c_{g,0} = 4.54 \times 10^6$ cm/s), radiation temperature $T_r$ (in unit of $1.67 \times 10^5$K) and opacity $\kappa$ (in unit of $0.34$ cm$^2$/g). The dashed blue lines indicate the location where the time averaged optical depth for the Rosseland mean opacity to the outer boundary of the simulation domain is unity.

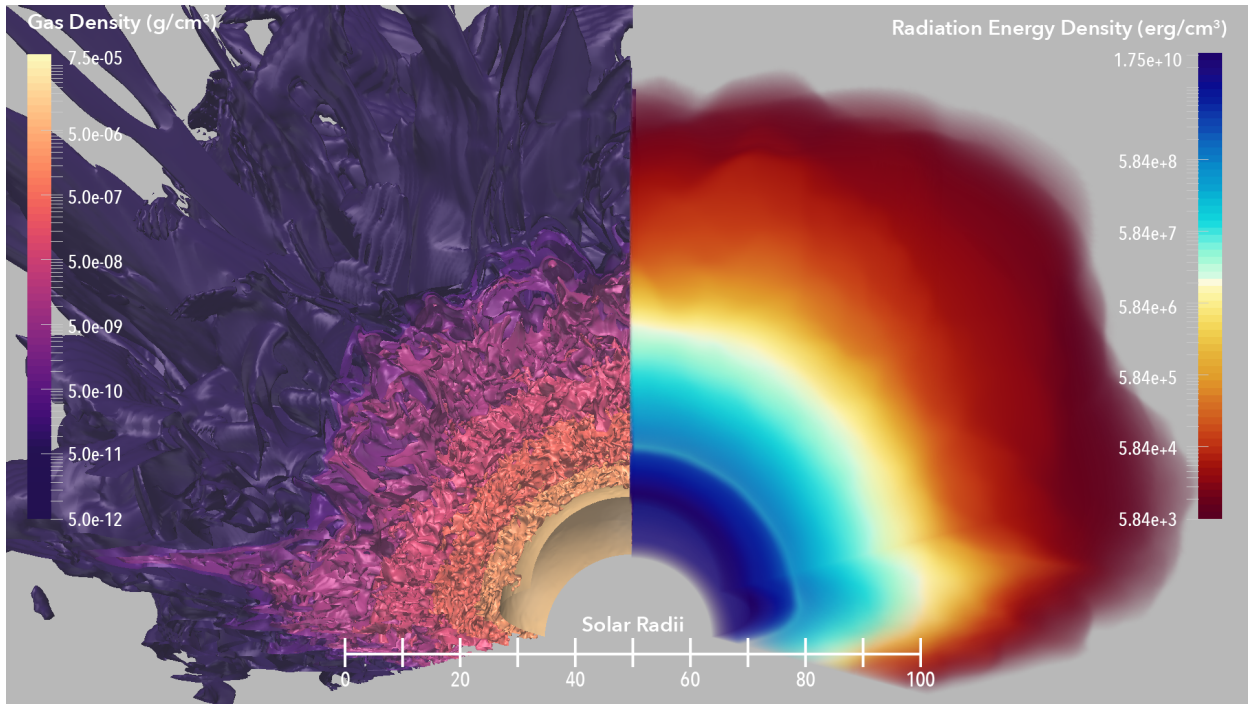

**Figure 3** | A snapshot of 3D density (left half panel) and radiation energy density (right half panel) for the run T19L6.4. The radial range covers $14.8R_\odot$ at the bottom to $336.5R_\odot$ at the top while the averaged photosphere location is at $102R_\odot$. Convection develops at the bottom due to the iron opacity peak at $44.6R_\odot$. The photosphere shows large scale plumes, which also cause strong variations of the radiation temperature at the photosphere across the surface of the star.

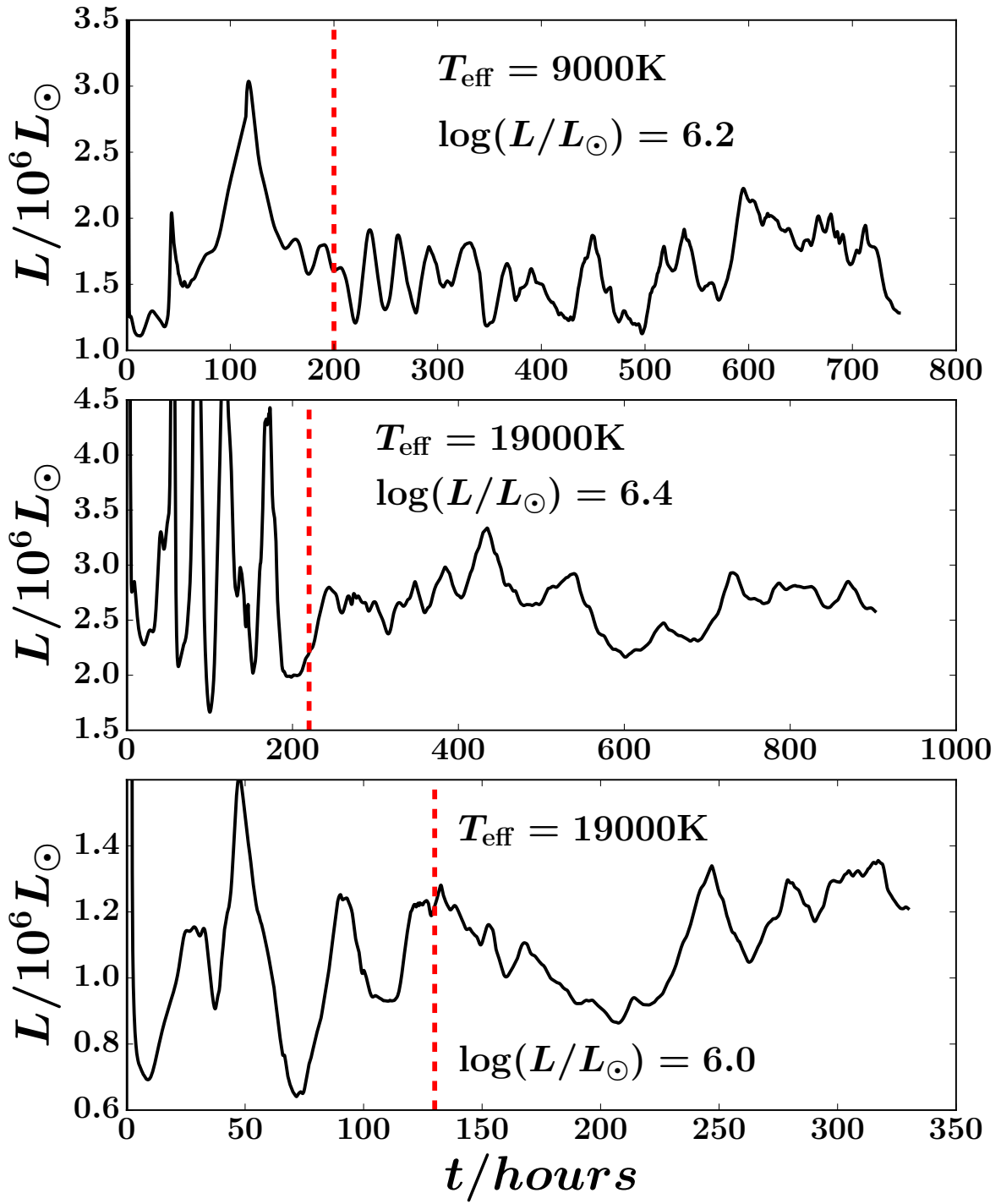

**Figure 4** | History of the total luminosity measured from the outer boundary of the simulation box

for the three simulations `T9L6.2` (top), `T19L6.4` (middle) and `T19L6` (bottom). The vertical dashed red lines indicate the time at which the runs reach steady state. The initial mass, effective temperature and averaged luminosity during steady state for the three simulations are labeled in each panel.

**Acknowledgements** We thank Joseph Insley from ALCF for helping us make the image shown in Figure 3, Nathan Smith for providing the data of LBVs and Bill Paxton for many conversations. This research was supported in part by the National Science Foundation under Grant No. NSF PHY 11-25915, 17-48958, and in part by a Simons Investigator award from the Simons Foundation (EQ) and the Gordon and Betty Moore Foundation through Grant GBMF5076. An award of computer time was provided by the Innovative and Novel Computational Impact on Theory and Experiment (INCITE) program. This research used resources of the Argonne Leadership Computing Facility and National Energy Research Scientific Computing Center, which are DOE Offices of Science User Facility supported under Contract DE-AC02-06CH11357 and DE-AC02-05CH11231. Resources supporting this work were also provided by the NASA High-End Computing (HEC) Program through the NASA Advanced Supercomputing (NAS) Division at Ames Research Center. The Flatiron Institute is supported by the Simons Foundation.

**Author Contributions**   Y.F.J. ran the simulations, analyzed the results and wrote the first draft of the paper. M.C. ran the MESA 1D stellar evolution calculations and made Figure 1. M.C., L.B., E.Q., O.B. and J.S. all read and commented on the draft.

**Competing Interests**   The authors declare that they have no competing financial interests.

**Correspondence**   Correspondence and requests for materials should be addressed to Yan-Fei Jiang (yanfei@kitp.ucsb.edu).